\documentclass[aps,superscriptaddress,showkeys,showpacs,twocolumn]{revtex4}
\usepackage{graphicx}
\newcommand{\ket}[1]{\vert #1 \rangle}
\newcommand{\bra}[1]{\langle #1 \vert}

\newcommand{\media}[1]{\langle #1 \rangle}

\begin{document}
\title{Teleportation improvement by inconclusive photon subtraction}
\author{Stefano Olivares}
\affiliation{Dipartimento di Fisica and Unit\'a INFM, Universit\`a degli
Studi di Milano, Italy}
\author{Matteo G. A. Paris}
\affiliation{Quantum Optics \& Information Group, INFM Unit\`a di
Pavia, Italy}
\author{Rodolfo Bonifacio}
\affiliation{Dipartimento di Fisica and Unit\'a INFM, Universit\`a degli
Studi di Milano, Italy}
\begin{abstract}
Inconclusive photon subtraction (IPS) is a conditional measurement
scheme to force nonlinear evolution of a given state. In IPS the
input state is mixed with the vacuum in a beam splitter and then
the reflected beam is revealed by ON/OFF photodetection. When the
detector clicks we have the (inconclusive) photon subtracted
state.  We show that IPS on both channels of an entangled
twin-beam of radiation improves the fidelity of coherent state
teleportation if the energy of the incoming twin-beam is below a
certain threshold, which depends on the beam splitter
transmissivity and the quantum efficiency of photodetectors. We
show that the energy threshold diverges when the transmissivity
and the efficiency approach unit and compare our results with that
of previous works on {\em conclusive} photon subtraction.
\end{abstract}
\maketitle
\section{Introduction}\label{s:intro}
Quantum information processing exploits quantum properties of physical
objects, basically entanglement, to improve performances of communication
channels and computational schemes. Perhaps, the most impressive quantum
information protocol realized so far is teleportation, where the inherent
nonlocality of entangled states is manifestly demonstrated. Teleportation
experiments have been mostly performed in the optical domain both for
polarization qubit \cite{tdm,zei} and coherent states of a continuous
variable (CV) system (a single mode radiation field) \cite{kim}. In
optical CV teleportation entanglement is provided by the so-called twin-beam
(TWB) of radiation
\begin{equation}\label{twb}
  \ket{{\rm twb}}\rangle_{ab} = \sqrt{1-x^2}\sum_{n=0}^{\infty} x^n
  \ket{n}_a \ket{n}_b,
\end{equation}
$a$ and $b$ being two modes of the field and $x$ the TWB
parameter (without loss of generality we can take $x$ real,
$0<x<1$). TWBs are produced by spontaneous downconversion in
nondegenerate optical parametric amplifiers. $\ket{{\rm
twb}}\rangle_{ab}$ is a pure state and thus its entanglement can
be quantified by the excess Von-Neumann entropy
\cite{bp91,lind,bp89,ve97}. The entropy of a two-mode state
$\varrho$ is defined as $S[\varrho] = - \hbox{Tr}\left\{\varrho
\log\varrho\right\}$ whereas the entropies of the two modes $a$
and $b$ are given by $S[\varrho_j] = -\hbox{Tr}_j\left\{\varrho_j
\log\varrho_j\right\}$, $j=a,b$, $\varrho_a =
\hbox{Tr}_b\left\{\varrho\right\}$ and $\varrho_b =
\hbox{Tr}_a\left\{\varrho\right\}$ denoting the partial traces.
The degree of entanglement of the state $\varrho$ is given by
$\Delta = S[\varrho_a] + S[\varrho_b] - S[\varrho]$, which
formalizes the idea that the stronger are the correlations in the
two-mode pure state, the more disordered should be the two modes
taken separately. Since $\ket{{\rm twb}}\rangle_{ab}$ is a pure
state, we have that $S[\varrho]=0$ and $S[\varrho_a]=S[\varrho_b]$
\cite{arak}, so that $\Delta=-\log(1-|x|^2) -|x|^2\log
|x|^2/(1-|x|^2)$ in terms of the TWB parameter and $\Delta=\log
(1+N/2) + N/2 \log (1+2/N)$ in terms of the number of photons of
the TWB $N=2 |x|^2/(1-|x|^2)$.  Notice that for \emph{pure} state
$\Delta$ represents the unique measure of entanglement \cite{po97}
and that TWBs are maximally entangled states for a given (average)
number of photons. The degree of entanglement is a monotonically
increasing function of either $|x|$ or $N$. The larger is the
entanglement the higher (closer to unit) is the fidelity of
teleportation based on TWB.
\par
The TWB parameter, which is sometimes also referred to as the
squeezing parameter of $|{\rm twb}\rangle\rangle_{ab}$, is given
by $x=\tanh {\cal G}$ with ${\cal G} \propto \chi^{(2)}L$,
$\chi^{(2)}$ being the nonlinear susceptibility of the crystal
used as amplifying medium and $L$ an effective interaction length.
For a given amplifier, the TWB parameter and thus the amount of
entanglement, is in principle fixed. However, since nonlinearities
are small, and the crystal length cannot be increase at will, it
is of interest to devise suitable quantum operations to increase
entanglement and in turn to improve teleportation fidelity. In
Ref. \cite{wel} Opatrn\'y {\em et al} suggest photon subtraction (PS)
as a scheme to increase entanglement, whereas in Ref. \cite{mil}
Milburn {\em et al} show that the fidelity of coherent state
teleportation with PS is indeed increased for any value of the
initial TWB parameter and of the coherent amplitude. PS is based
on mixing each channel of a TWB with the vacuum in a high
transmissivity beam splitter and then counting the number of
photons in one of the outgoing arms (probe mode). Upon detecting
exactly one photon in both probes one has the photon-subtracted
and thus  entanglement enhanced, state to be used in a teleporting
device. Although recent developments toward effective
photocounting are encouraging, the discrimination of one photon
from zero, two, three and so on is still experimentally
challenging. Therefore, the PS scheme studied in Refs.
\cite{wel,mil} appears to be of difficult implementation. Here we
focus our attention on the experimental implementation of CV
teleportation and, therefore, analyze a different PS scheme where
photodetection after the beam splitter is inconclusive {\em i.e.}
it is performed by a realistic ON/OFF (Geiger-like) detector which
only discriminates, with quantum efficiency $\eta$, the vacuum
from the presence of any number of photons. As we will see,
inconclusive photon subtraction (IPS) is an effective method to
increase photon correlations of TWB and, in turn, to improve
fidelity in coherent state teleportation, provided that the
initial TWB energy is below a certain value.
\par
The paper is structured as follows: in Section \ref{s:ips} we
describe in details the IPS scheme, whereas in Section \ref{s:tel}
we analyze the use of the IPS output state in coherent state
teleportation.  Section \ref{s:outro} closes the paper with some
concluding remarks.
\section{The inconclusive photon subtraction scheme}
\label{s:ips}
\begin{figure}[h!]
\includegraphics[width=0.4\textwidth]{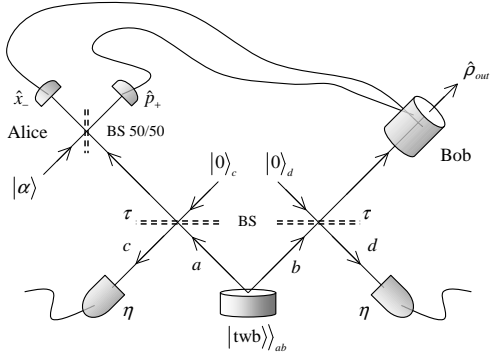}
\caption{Schematic diagram of continuous variable optical quantum
teleportation assisted by inconclusive photon subtraction.}\label{f:scheme}
\end{figure}

\begin{figure}[h!]
\includegraphics[width=0.4\textwidth]{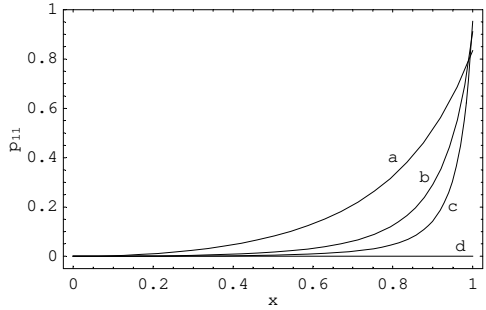}
\caption{Conditional probability $p_{11}(x,\tau_{\hbox{eff}})$
of obtaining the IPS state as a function of
the TWB parameter $x$ for different values of the effective
transmissivity $\tau_{\rm eff}=0.5$ (a), $0.8$ (b), $0.9$ (c)
and $1$ (d).}\label{f:p11}
\end{figure}

In IPS the two channels of TWB impinge onto two beam splitters each with
transmissivity $\tau$, which we consider equal, where they are mixed
with the vacuum state $\ket{0}_c\ket{0}_d$ of modes $c$ and $d$
(see Figure \ref{f:scheme}). The effect of the beam splitter on two modes,
say $a$ and $c$, is described by the unitary operator
\begin{equation}\label{bs}
{U}_{ac}(\tau) = \exp\left[ \lambda_\tau ({a} {c}^\dag - {a}^\dag {c})
\right],
\end{equation}
where
\begin{equation}\label{lambda:tau}
  \lambda_\tau = \arctan \left( \sqrt{\frac{1-\tau}{\tau}} \right)
\end{equation}
and ${a}$, ${c}$, ${a}^\dag$ and ${c}^\dag$ are
the annihilation and creation operators for modes $a$ and $c$,
respectively. After the beam splitters the wave function of the
system is
\begin{widetext}
\begin{eqnarray}\label{psi:bs}
  \ket {\psi_{\rm BS}} &=& {U}_{ac}(\tau) {U}_{bd}(\tau) \, \ket{{\rm twb}}\rangle_{ab}\ket{0}_c
  \ket{0}_d \nonumber \\
  &=& \sqrt{1-x^2} \sum_{n=0}^{\infty}
  (x \tau)^{n} \sum_{p,q=0}^{n} \left( \frac{1-\tau}{\tau}
  \right)^{\frac{p+q}{2}} \sqrt{ {n \choose p}{n \choose q} }
  \ket{n-p}_a \ket{n-q}_b \ket{p}_c \ket{q}_d.
\end{eqnarray}
\end{widetext}
Now, we perform a conditional \emph{inconclusive photon
subtraction} revealing the mode $c$ and $d$ by ON/OFF
photodetection. The POVM $\{\Pi_0(\eta),\Pi_1(\eta)\}$
(positive operator-valued measure) of each ON/OFF detector
is given by
\begin{equation}\label{povm1}
  {\Pi}_0 (\eta) = \sum_{j=0}^{\infty} (1-\eta)^j
  \ket{j}\bra{j}\:,
\qquad   {\Pi}_1(\eta) = {\openone} - {\Pi}_0 (\eta)\:,
\end{equation}
$\eta$ being the quantum efficiency. Overall, the conditional measurement
on the modes $c$ and $d$, is described by the POVM
\begin{eqnarray}
{\Pi}_{00} (\eta)&=&{\Pi}_{0,c} (\eta) \otimes {\Pi}_{0,d}
(\eta)\:, \\ {\Pi}_{01} (\eta) &=& {\Pi}_{0,c} (\eta) \otimes
{\Pi}_{1,d} (\eta)
\\ {\Pi}_{10} (\eta)&=&{\Pi}_{1,c} (\eta) \otimes {\Pi}_{0,d}
(\eta)\:, \\ {\Pi}_{11} (\eta)&=&{\Pi}_{1,c} (\eta) \otimes
{\Pi}_{1,d} (\eta) \label{povm11}\;.
\end{eqnarray}
We are interested in the situation when both the detectors click.
The corresponding conditional state for the modes $a$ and $b$ will
be referred to as the IPS state. Notice that this kind of measurement
is \emph{inconclusive}, {\em i.e.} it does not discriminate the number of photons
present in the beams: one can only say that a certain unknown
number of photons has been revealed and, then, subtracted from
each mode and that this number, in general, is not the same for the
two modes.
The probability of observing a click in both the detectors is given by
\begin{widetext}
\begin{eqnarray}\label{p11}
  p_{11}(x,\tau,\eta) &=& \hbox{Tr}_{abcd}
 \left\{ {\varrho}_{\rm BS}\:{\openone}_{a} \otimes {\openone}_{b}
 \otimes \Pi_{11} (\eta) \right\}\nonumber   \\
  &=&\frac{x^2 \eta^2 (1-\tau)^2 \left\{ 1+x^2 [ 1 - \eta ( 1-\tau )]
  \right\}}{\left\{ 1-x^2 [ 1 - \eta ( 1-\tau )] \right\}
  \left\{ 1-x^2 [ 1 - \eta ( 1-\tau )]^2 \right\}}
\:, \end{eqnarray}
where ${\varrho}_{\rm BS} = \ket {\psi_{\rm
BS}}\bra{\psi_{\rm BS}}$ and the corresponding conditional state
reads as follows
\begin{eqnarray}
\label{ips}
{\varrho}_{\rm IPS}(x,\tau,\eta) &=&
\frac{ {\rm Tr}_{cd} \left\{ {\varrho}_{\rm BS} \, {\openone}_{a} \otimes
{\openone}_{b} \otimes {\Pi}_{11}(\eta) \right\} }{p_{11}(x,\tau,\eta)}
  \nonumber \\ &=& \frac{1-x^2}{p_{11}(x,\tau,\eta)}
  \sum_{n,m=0}^{\infty} (x \tau)^{n+m} \sum_{h,k=0}^{{\rm Min}[n,m]} f_{h,k}(\tau,\eta)
  \sqrt{ {n \choose h}{n \choose k}{m \choose h}{m \choose k} }
  \nonumber \\
  &\mbox{}& \hspace{5cm} \times \ket{n-k}_a \ket{n-h}_b {_b}\bra{m-h} {_a}\bra{m-k}
\end{eqnarray}
where
\begin{equation}\label{ips:fhk}
  f_{h,k}(\tau,\eta)= \left[ 1 - (1-\eta)^h \right] \left[ 1 - (1-\eta)^k
  \right] \left( \frac{1-\tau}{\tau} \right)^{h+k}\:.
\end{equation}
\end{widetext}
The mixing with the vacuum in a beam splitter with transmissivity
$\tau$ followed by ON/OFF detection with quantum efficiency $\eta$ is
equivalent to mixing with an effective transmissivity
\begin{equation}\label{tau:eff}
  \tau_{\rm eff} (\tau,\eta) = 1 - \eta(1-\tau)
\end{equation}
followed by an ideal (\emph{i.e.} efficiency equal to one) ON/OFF
detection. Therefore, the IPS state (\ref{ips}) can be studied
for $\eta = 1$ and replacing $\tau$ with $\tau_{\rm eff}$. In
this way, the conditional probability (\ref{p11}) of obtaining the
IPS state rewrites as
\begin{equation}\label{p11:eff}
  p_{11}(x,\tau_{\rm eff})=\frac{x^2 (1-\tau_{\rm eff})^2
  (1+x^2 \tau_{\rm eff})}{(1-x^2 \tau_{\rm eff})(1-x^2\tau_{\rm
  eff}^2)},
\end{equation}
which, in general, is larger than the corresponding probability
for \emph{conclusive} photo-subtraction methods, where the same
(known) number of photons is subtracted from the input states
\cite{wel,mil} (Figure \ref{f:p11}). In fact, in IPS case the
coincidence between the two detectors can occur also when a
different (unknown) number of photons is revealed.
\par
\begin{figure}[h!]
\includegraphics[width=0.4\textwidth]{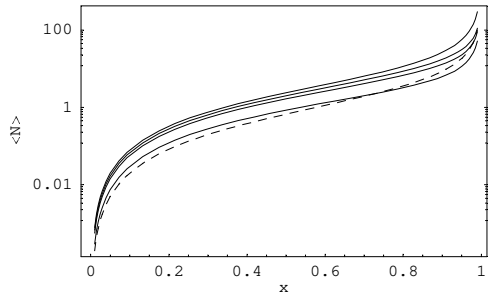}
\caption{Log-Linear plot of TWB (dashed line) and IPS state
average photon number as a function of the TWB parameter for
different values of $\tau_{\rm eff} = 1 - \eta(1-\tau)$
(solid lines from top to bottom: $\tau_{\rm eff}=1$, $0.9$, $0.8$
and $0.5$). }\label{f:avn}
\end{figure}

In Figure \ref{f:avn} we plot the average photon number of TWB and
of the IPS state: below a certain threshold value for $x$ the
energy of the IPS state is increased. As a matter of fact, the IPS
state is no longer a pure state and, therefore, the excess
Von-Neumann entropy cannot be used to quantify the degree of
entanglement. In order to characterize the IPS state we analyze
the quantity
\begin{widetext}
\begin{equation}\label{delta:n}
  \Delta_{a,b}(x,\tau_{\rm
  eff})=\frac{\media{{d}^2}-\media{{d}}^2}{\media{{n}_a+{n}_b}}=
  \frac{(1-\tau_{\rm eff})(1-x^2\tau_{\rm eff}^2)^2(2-x^2-x^4\tau_{\rm
eff})}{(1+\tau_{\rm eff})(1-x^2\tau_{\rm eff})[2-x^2(1+\tau_{\rm eff}+
\tau_{\rm eff}^2)+x^6\tau_{\rm eff}^3]}
\end{equation}
\end{widetext}
where ${d}={n}_a - {n}_b$ with ${n}_a = {a}^\dag {a}$
and ${n}_b = {b}^\dag {b}$. $\Delta_{a,b}(x,\tau_{\rm eff})=$ in
Eq.  (\ref{delta:n}) is a measure of
the \emph{difference number squeezing} {\em i.e.} of the photon correlation
between the two modes rather than entanglement. Notice that for TWB
$\Delta_{a,b}(x) = 0$, \emph{i.e.}
this state shows perfect correlation in the photon number. For $\tau_{\rm eff}$
approaching unit, Eq. (\ref{delta:n}) can be
approximated by $\Delta_{a,b}(x,\tau_{\rm eff} \rightarrow 1)
\approx \frac12 (1-\tau_{\rm eff})$
\emph{i.e.} $\Delta_{a,b}(x,\tau_{\rm eff})$
becomes independent from the TWB parameter $x$.
As we will see in the next Section, although the photon correlation
in the IPS state is apparently decreased, the fidelity of IPS-based
teleportation is increased with respect to TWB.
\section{Inconclusive photon subtraction and teleportation}
\label{s:tel} In order to implement quantum teleportation, the IPS
state (\ref{ips}) is shared between Alice and Bob (Figure
\ref{f:scheme}). Alice mixes the mode $a$ of the IPS state with a
given quantum state, which she wishes to teleport to Bob, on a
50/50 beam-splitter and then she measures the two conjugated
quadratures $x_{-}=\frac12(e+e^\dag)$ and $p_{+}=\frac{i}2(f^\dag
-f)$ (corresponding to position difference and momentum sum
respectively), $e$ and $f$ being the two modes outgoing the beam
splitter. These results are classically sent to Bob, who applies a
displacement by the amount $\beta = x_{-} + i p_{+}$ to his mode
$b$. If $\sigma$ is the density matrix of the state to be
teleported, the measurement performed by Alice is equivalent to a
generalized heterodyne detection, described by the following POVM
(acting on mode $a$) \cite{het}
\begin{equation}\label{POVM:tele}
{\Pi}_{a}(\beta) = \frac{1}{\pi}
{D}(\beta) \, {\sigma}^{T} \, {D}^{\dag}(\beta)
\end{equation}
where $\beta$ is a complex number, ${D}(\beta)=\exp \{ \beta
{a}^{\dag} - \beta^{*} {a} \}$ is the displacement
operator, $(\cdots)^{T}$ stands for the transposition operation.
The probability for the outcome $\beta$ is
\begin{equation}\label{tele:prob}
p(x,\tau,\eta,\beta) = {\rm Tr}_{ab} \left\{ {\varrho}_{\rm
IPS} \, {\Pi}_{a}(\beta) \otimes {\openone}_{b} \right\},
\end{equation}
and the conditional state is
\begin{equation}\label{rho:b}
  {\varrho}_{b}(x,\tau,\eta,\beta) = \frac{{\rm Tr}_{a}
  \left\{ {\varrho}_{\rm IPS} \, {\Pi}_{a}(\beta) \otimes
  {\openone}_{b}\right\}}{p(x,\tau,\eta,\beta)},
\end{equation}
which, after the displacement by Bob, becomes ${\varrho}_{\rm out}(x,\tau,\eta,
\beta) = {D}^{\dag}(\beta) \, {\varrho}_{b} \, {D}(\beta)$. \par
\begin{figure}[htb]
\includegraphics[width=0.4\textwidth]{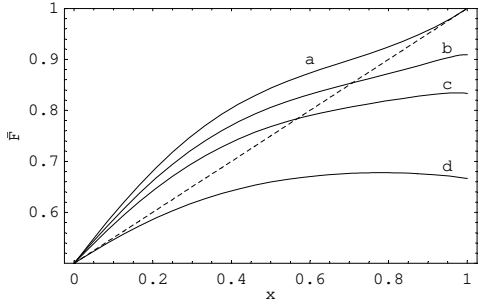}
\caption{IPS average fidelity $\overline{F}(x,\tau_{\rm eff})$
as a function of the TWB parameter
for different values of $\tau_{\rm eff} = 1 - \eta(1-\tau)$
($\tau_{\rm eff}=1$ (a), $0.9$ (b), $0.8$ (c) and $0.5$ (d));
the dashed line is the average fidelity $\overline{F}_{\rm TWB}(x)$
for teleportation with TWB.}\label{f:fidelity}
\end{figure}
\begin{figure}[htb]
\includegraphics[width=0.4\textwidth]{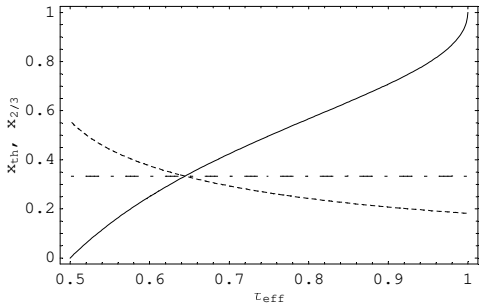}
\caption{Threshold value $x_{\rm th}(\tau_{\rm eff})$ on
the TWB parameter $x$ (solid line): when $x < x_{\rm th}$ we have
$\overline{F}(x,\tau_{\rm eff}) > \overline{F}_{\rm TWB}(x)$ and teleportation
is improved. The dot-dashed line is $x=1/3$, which corresponds to
$\overline{F}_{\rm TWB}=2/3$: when fidelity is greater than 2/3 Bob is sure
that his teleported state is the best existing copy of the initial state
\cite{gross}.  The dashed line represents the values $x_{2/3}(\tau_{\rm eff})$
giving an average fidelity $\overline{F}(x,\tau_{\rm eff})=2/3$.
When $x_{2/3}<x<x_{\rm th}$ both
the teleportation is improved and the fidelity is greater than
$2/3$.}\label{f:xvstaueff}
\end{figure}

For coherent state teleportation $\sigma=|\alpha\rangle\langle\alpha |$
we have
\begin{widetext}
\begin{eqnarray}\label{rho:out}
{\varrho}_{\rm out}(x,\tau,\eta,\beta) &=&
\frac{1-x^2}{p_{11}(x,\tau,\eta) \, p(x,\tau,\eta,\beta)}
\sum_{n,m=0}^{\infty} (x \tau)^{n+m} \sum_{h,k=0}^{{\rm Min}[n,m]}
f_{h,k}(\tau,\eta) \nonumber\\
&\mbox{}&\hspace{0.2cm} \times \sqrt{ {n \choose h}{n \choose k}{m
\choose h}{m \choose k} } \, \frac{e^{-|\alpha + \beta|^2}
(\alpha^{*} + \beta^{*})^{n-k} (\alpha + \beta)^{m-k}}{\sqrt{(n-k)! \, (m-k)!}} \nonumber\\
&\mbox{}& \hspace{7.5cm} \times \ket{m-h}_b {_b}\bra{n-h}\:,
\end{eqnarray}
\end{widetext}
where $p_{11}(x,\tau,\eta)$, $f_{h,k}(\tau,\eta)$ and
$p(x,\tau,\eta,\beta)$ are given by Eqs. (\ref{p11}),
(\ref{ips:fhk}) and (\ref{tele:prob}) respectively. Therefore the
teleportation fidelity, $F(x,\tau,\eta,\beta)$, and the average
fidelity, $\overline{F}(x,\tau,\eta)$, are given by
\begin{widetext}
\begin{eqnarray}\label{tele:fid:beta}
  F(x,\tau,\eta,\beta) &\equiv& \bra{\alpha}{\varrho}_{\rm out} \ket{\alpha}
  \nonumber\\
  &=& \frac{1-x^2}{p_{11}(x,\tau,\eta) \, p(x,\tau,\eta,\beta)}
  \sum_{n,m=0}^{\infty} (x \tau)^{n+m} \sum_{h,k=0}^{{\rm Min}[n,m]} f_{h,k}(\tau,\eta)
  \nonumber\\
  &\mbox{}& \hspace{.5cm} \times \sqrt{ {n \choose h}{n \choose k}{m \choose h}{m \choose k} }
  \, \frac{e^{-2|\alpha + \beta|^2} |\alpha + \beta|^{2(m+m-h-k)}}{\sqrt{(n-k)! \, (m-k)! \, (n-h)! \,
  (m-h)!}}
\end{eqnarray}
and
\begin{eqnarray}\label{tele:fid:ave}
  \overline{F}(x,\tau,\eta) &\equiv& \int {\rm d}^2 \beta \, p(x,\tau,\eta,
  \beta)F(x,\tau,\eta,\beta)
  \nonumber\\
  &=& \frac{1-x^2}{2 \, p_{11}(x,\tau,\eta)}
  \sum_{n,m=0}^{\infty}
  \left( \frac{x \tau}{2} \right)^{n+m} \sum_{h,k=0}^{{\rm Min}[n,m]} 2^{h+k} f_{h,k}(\tau,\eta)
  \nonumber\\
  &\mbox{}& \hspace{1cm} \times \sqrt{ {n \choose h}{n
  \choose k}{m \choose h}{m \choose k} } \, \frac{(n+m-h-k)!}{\sqrt{(n-h)! \, (n-k)! \, (m-h)! \,
  (m-k)!}}\:.
\end{eqnarray}
\end{widetext}
By the substitution $\eta \rightarrow 1$ and $\tau \rightarrow
\tau_{\rm eff}=1-\eta(1-\tau)$ Eq. (\ref{tele:fid:ave}) can be
summed, leading to the following expression
\begin{widetext}
\begin{eqnarray}\label{ave:summed}
  \overline{F}(x,\tau_{\rm eff}) = \frac12
  \frac{(1+x)(1+x \tau_{\rm eff})(1-x^2\tau_{\rm eff})}{(1+x^2
  \tau_{\rm eff})[1+(1-\tau_{\rm eff})x]}
  \frac{[2-2x \tau_{\rm eff} + x^2
  \tau_{\rm eff}]}{\{ 2 - [2+(1-\tau_{\rm eff})x]x \tau_{\rm eff}
  \}}\:.
\end{eqnarray}
\end{widetext}
In Figure \ref{f:fidelity} we plot the average fidelity for
different values of $\tau_{\rm eff}$: the IPS state improves the
average fidelity of quantum teleportation when the energy of the
incoming TWB is below a certain threshold, which depends on
$\tau_{\rm eff}$ and, in turn, on $\tau$ and $\eta$ (see Eq.
(\ref{tau:eff})). When $\tau_{\rm eff}$ approaches unit
(when $\eta \rightarrow 1$ and $\tau \rightarrow 1$), Eq.
(\ref{ave:summed}) reduces to the result obtained by Milburn
\emph{et al} in Ref. \cite{mil} and the IPS average fidelity (line
labelled with ``a'' in Figure \ref{f:fidelity}) is always greater
than the one obtained with the TWB state (\ref{twb}), \emph{i.e.}
\begin{equation}\label{twb:fid}
  \overline{F}_{\rm TWB}(x) = \frac{1+x}{2}.
\end{equation}
However, a threshold value, $x_{\rm th}(\tau_{\rm eff})$, for the
TWB parameter $x$ appears when $\tau_{\rm eff} < 1$: only if
$x$ is below this threshold the teleportation is actually improved
($\overline{F}(x,\tau_{\rm eff}) > \overline{F}_{\rm TWB}(x)$), as
shown in Figure \ref{f:xvstaueff}. Notice that, for $\tau_{\rm eff}
< 0.5$, $\overline{F}(x,\tau_{\rm eff})$ is always below
$\overline{F}_{\rm TWB}(x)$.
\par
A fidelity larger than $1/2$ is needed to show that a truly 
nonlocal information transfer occurred \cite{tbr}. Notice 
that using both the TWB (\ref{twb}) and the IPS state (\ref{ips}), 
this limit is always reached (Figure \ref{f:fidelity}). 
Nevertheless, we remember
that in teleportation protocol the state to be teleported is
destroyed during the measurement process performed by Alice, so
that the only remaining \emph{copy} is that obtained by Bob. When
the initial state carries reserved information, it is
important that the only existing copy  will be the Bob's one. On
the other hand, using the usual teleportation scheme, Bob cannot
avoid the presence of an eavesdropper, which can clone the state,
obviously introducing some error \cite{cerf}, but he is able to to
verify if his state was duplicated. This is possible by the
analysis of the average teleportation fidelity: when fidelity is
greater than $2/3$ \cite{ralph}, Bob is sure that his state was not cloned
\cite{cerf,gross}. The dashed line in Figure \ref{f:xvstaueff} shows
the values $x_{2/3}(\tau_{\rm eff})$ which give an average
fidelity (\ref{ave:summed}) equal to $2/3$: notice that when
$x_{2/3} < x < x_{\rm th}$ both the teleportation is improved and
the the fidelity is greater than $2/3$. Moreover, while the
condition $\overline{F}_{\rm TWB}(x) > 2/3$ is satisfied only if
$x > 1/3$, for the IPS state there exists a $\tau_{\rm
eff}$-dependent interval of $x$ values ($x_{\rm 2/3} < x < 1/3$)
for which teleportation can be considered \emph{secure}
($\overline{F}(x,\tau_{\rm eff})
> 2/3$). 
\section{Conclusions}\label{s:outro}
We have analyzed a photon subtraction scheme similar to that of
Refs. \cite{wel,mil} to modify twin-beam and improve coherent
state teleportation. The difference in our analysis is that the
conditional photodetection after the beam splitters is considered
inconclusive, {\em i.e.} performed by ON/OFF detectors which do
not discriminate among the number of photons. This is closer to
the current experimental situations and provides a higher
conditional probability. We found that fidelity is improved
compared to that of TWB-based teleportation if the initial TWB
parameter is smaller than a threshold value, which in turn depends
on the beam splitter transmissivity and on the quantum efficiency
of the photodetectors. For realistic values of these parameters
($\eta$ larger than $90\%$ and $\tau$ larger than $99\%$) the
threshold is close to unit. In addition, there exists a interval
of $x$ for which teleportation can be considered \emph{secure},
{\em i.e} the receiver is able to check whether or not the state
has been duplicated before teleportation. We conclude that IPS on
TWB is a robust and realistic scheme to improve coherent state
teleportation using current technology.
\section*{Acknowledgments} 
This work has been sponsored by the INFM through 
the project PRA-2002-CLON, by MIUR through the PRIN projects {\em Decoherence
control in quantum information processing} and {\em Entanglement assisted high
precision measurements}, and by EEC through the project IST-2000-29681 (ATESIT).
MGAP is research fellow at {\em Collegio Alessandro Volta}.

\end{document}